\documentclass[mathleft]{an}
\usepackage{graphicx}
\usepackage{times}
\usepackage{amssymb}
\begin{document}
\newcommand{\dd}{deg$^{2}$}
\newcommand{\flux}{$\rm erg \, s^{-1} \, cm^{-2}$}
\newcommand{\LL}{$\lambda$}

% The following seven commands are intended for editorial usage and should be ignored by
% the author(s).
\Pagespan{1}{}% Document's page range. 
% If second parameter is left empty, the last page is computed automatically.
\Yearpublication{2006}%
\Yearsubmission{2005}%
\Month{11}%   
\Volume{999}%  
\Issue{88}% 
% \DOI{This.is/not.aDOI}% 

\title{The XXL survey: first results and future}

\author{M. Pierre$^{1}$\fnmsep\thanks{Corresponding author:
\email{mpierre@cea.fr}\newline}
C. Adami$^{2}$, M. Birkinshaw$^{3}$, L. Chiappetti$^{4}$, S. Ettori$^{5}$, A. Evrard$^{6}$, L. Faccioli$^{1}$, F. Gastaldello$^{4}$, P. Giles$^{3}$, C. Horellou$^{7}$
% ...
}

\institute{The affiliations of the 111 authors are listed in section 6}
\titlerunning{XXL and XXL-II}
\authorrunning{M. Pierre et al}

\received{***}
\accepted{***}
%\publonline{later}


\abstract{The XXL survey currently covers two 25 \dd\ patches with XMM observations of $\sim$ 10ks. We summarise the scientific results associated with the first release of the XXL data set, that occurred mid 2016. We review several arguments for increasing the survey depth to 40 ks during the next decade of XMM operations. X-ray ($z<2$) cluster, ($z<4$) AGN and cosmic background survey science will then benefit from an extraordinary data reservoir. This, combined with deep multi-\LL\ observations, will lead to solid standalone cosmological constraints and provide a wealth of information on the formation and evolution of AGN, clusters and the X-ray background. In particular,  it will offer a unique opportunity to pinpoint the $z>1$ cluster density. It will eventually  constitute a reference study and an ideal calibration field for the upcoming eROSITA and Euclid missions.}

\maketitle

\section{Introduction}

Almost 17 years after the launch of {\sc XMM-Newton}, it is timely  to review its scientific achievements. A thorough census of the still open or newly  raised questions will help us to optimise the use of the observatory for its last decade. In this paper, we focus on medium-deep extragalactic surveys. More specifically, we scrutinise the contribution of X-ray large-scale structure studies to the global multi-\LL\ and multi-probe effort toward precision cosmology.  Although XMM was not initially designed as a survey instrument, its large field of view, good PSF and unrivalled collecting area provide a unique opportunity to scan the structure of the energetic universe. The mosaic observing mode implemented in 2008 further enhanced these capabilities. \\
Starting from the Guaranteed Time pooled by the Li\`ege, Milano and Saclay groups at the very beginning of the XMM mission, we undertook a uniform mapping of the extragalactic sky. Subsequent Guest Observer observations of $\sim$ 10 ks allowed us to achieve a coverage of some 11 \dd\ by 2009 (Elyiv et al 2012, Chiappetti et al 2013, Clerc et al 2014). This XMM-LSS pilot survey was an essential  step in understanding the X-ray cluster selection function - down to a depth never explored to far  - and in testing its impact on the scaling relations and subsequent cosmological analysis.   In 2010, we were allocated an XMM Very Large Programme to extend the coverage to two areas of 25 \dd\ each at the same sensitivity: the XXL survey\footnote{http://irfu.cea.fr/xxl}. \\
The main driver of the XXL survey is cosmology, based  on both AGN and cluster counts along with 3D topological and environmental studies; hence the need for a large connected area, rather than serendipitous archival detections. Other fundamental motivations for promoting a large-scale uniform X-ray coverage include the  simplification of the selection function and the availability of a set of associated homogeneous surveys covering the entire electromagnetic spectrum on the same area (from UV to radio). This enables coherent source identification along with uniform SED and redshift measurements, which constitute the two fundamental steps toward the census of the  cluster and AGN populations and their characterisation.\\
In this paper, we first recall the main issues pertaining to cluster cosmology, then summarise the outcome of the recent series of XXL articles. In the last sections, we propose a route for extending the current existing data set and provide a truly outstanding scientific legacy. The articles from the first XXL series are  quoted with roman numbers in square brackets.

\section{Cluster cosmology and the motivations of the XXL project} 
\begin{figure}
%\begin{center}
\hspace{-0.5cm}
\includegraphics[width=8.75cm]{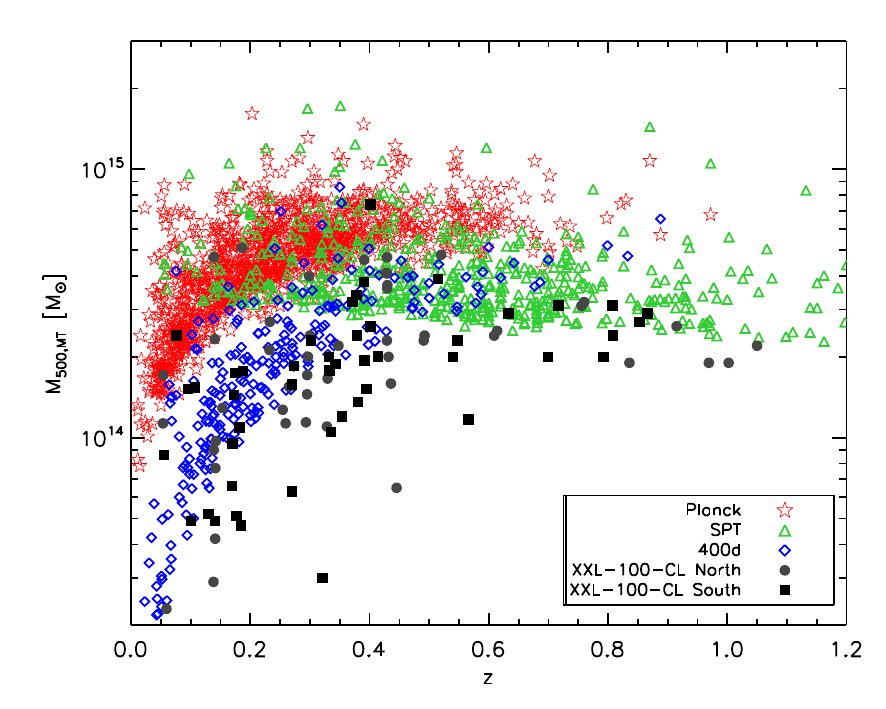}
\caption{Mass range covered by the XXL brightest 100 clusters, compared to other surveys. {\em Credit: Pacaud et al, A\&A 592, A2, 2016, reproduced with permission \copyright ~ ESO} }
%\end{center}
\label{mz}
\end{figure}

As the most massive self-gravitating entities of the universe, clusters of galaxies are theoretically key objects to constrain cosmological models: they are both sensitive to the geometry of the space-time and to structure growth. Originating from  physical processes different from those of the cosmic microwave background (CMB), supernovae and baryonic acoustic oscillations, they should to provide independent and complementary constraints. However, a number of practical difficulties, most of them having been overlooked before the advent of XMM and {\sc Chandra}, render such a study especially challenging. These include: 1) It is now well established that the X-ray selection function of these extended objects cannot be modelled by a simple flux limit but should be estimated in the flux-size parameter plane. 2) Scaling relations, that enable the use of mass proxies (e.g. Lx, Tx Mgas or the optical richness), are very much dependent on the samples on which they are based; disentangling the selection effects requires the knowledge of the intrinsic scatter of these relations; however, very few scatter measurements exist and most of the time, one relies on assumptions from numerical simulations. 3) The fact that cluster masses are not a direct observable continues to feed a lively controversy, motivating innovative observational studies ; to this should be added that hydrodynamical simulations indicate a bias up to 20-30\% between true and hydrostatic masses. 4) The whole picture must be consistently worked out in an evolving environment, while the evolution of the cluster baryonic physics is still very much debated. Rigorously, cosmology, cluster evolution and selection effects should be addressed in a self-consistent approach (for a review on these topics see e.g. Allen et al 2011).\\
In this context, the XXL survey aims at an independent and self-consistent cosmological analysis. As much as possible, scaling relations are derived from the cluster sample itself in conjunction with measurements in other wavebands like the integrated K-band luminosity or deep weak lensing information. The interplay between cluster and AGN physics as well as its impact on cluster detection and scaling relations is studied with great care via several sets of numerical simulations. Given its 50 \dd\ coverage, XXL tackles the very important, and still largely unexplored, $M_{500} \sim 5\times 10^{13}-2\times 10^{15}~ M_\odot$ regime for $z\sim 0.5$ clusters and thus provides information complementary to the {\sc Planck, SPT} and {\sl Weighing the Giants}  samples (Fig. \ref{mz}).

\section{First results from the XXL survey}

The XXL survey gathers some 100 scientists worldwide and is accompanied by a comprehensive multi-\LL\ and spectroscopic programme. The two surveyed 25 \dd\ areas  (XXL-N: RA = 2h30 Dec = -4d30' ; XXL-S: RA = 23h30 Dec = -55d00') are covered by more than 500 independent XMM observations totalling some 6.9 Ms, which makes  XXL the largest XMM programme to date. It was designed such as to provide a sample of some 500 clusters of galaxies out to a redshift of unity, suitable for cosmological study. The point-source sensitivity is $\sim 5 \times 10^{-15}$ \flux\ in the [0.5-2] keV band. The survey characteristics along with its  extensive imaging+spectroscopic associated follow-up and simulation programmes is presented in [paper I]. In June 2016, the first series of XXL results was published in a special issue of {\sl Astronomy and Astrophysics}\footnote{http://www.aanda.org/articles/aa/abs/2016/08/contents/contents.html}.
They are based on the brightest 100 clusters and 1000 AGN samples. Both X-ray catalogues along with two associated VLA and ATCA radio source lists are available at the CDS. They can also be retrieved in a more extensive form, along with the XMM images and exposure maps, via the XXL  databases (Table \ref{cat}).
The XXL team pays special attention to the delivery of well-validated catalogues and, beside the science publications, considers  this legacy aspect  a priority commitment.

\begin{table}
% \centering%%%
\caption{The XXL databases. As of the 2016 release, the X-ray catalogues are limited to the brightest 100 clusters and 1000 AGN. Incrementally deeper releases will follow. In addition, the database in Milano provides the X-ray raw and wavelet MOS+pn mosaic images as well as all exposure maps for both XXL fields up to AO11.  }
\label{cat}
\hspace{-0.5cm}
\begin{tabular}{ll}\hline \hline
%Content & Address\\ 
CLUSTERS & {\bf http://xmm-lss.in2p3.fr:8080/xxldb/index.html} \\
\em ~& X-ray and optical images\\
~& Details of the redshift calculations\\
~& X-ray: flux, luminosity, temperature\\
~& Mass estimate\\
\hline
AGN & {\bf http://cosmosdb.iasf-milano.inaf.it/XXL/}\\
~ & Fluxes, X-ray spectral fits, counterparts\\
~ &VLA 3GHz and ATCA 2.1GHz radio catalogues \\
~ & AAOmega redshifts\\  
\hline
\end{tabular}
\end{table}

\subsection{Summary of the first results}
The 2016 results pertain to about 1/5 and 1/20 of the complete cluster and AGN samples respectively. They already provide  interesting clues that can be summarised as follows:
\subsubsection{Clusters}
 1) We performed an internally consistent derivation of the M-T and L-T relations  [papers III, IV]; 2) The luminosity function does not show evolution out to a redshift of unity [paper II], while the  L-T relation is compatible with self-similar evolution [paper III]; 3) The  modelling of the cluster number counts shows a deficit with respect to predictions assuming the Planck CMB cosmology; 4) The low gas content of these clusters favours strong AGN feedback activity [paper XIII]; 5) We discovered five superclusters [papers II, VII]; 6) We have detected via the Sunyaev-Zel'dovich effect (S-Z) one of the XXL distant cluster candidates, which turned out to be the highest-redshift cluster ( $z\sim 1.9$ ) ever detected to-date in S-Z [papers V].  
\subsubsection{AGN}
1) We improved upon the photometric redshift determination for AGN by applying a Random Forest classification trained to identify  the optimal photometric redshift category for each object (passive, star forming, starburst, AGN, QSO); 2) The X-ray spectral properties are consistent with those of the bright sources from the literature; 3) The [2-10] keV luminosity function over the $0.01<z<3.0$ range favours the Luminosity Dependent Density Evolution model; 4) A large cluster of AGN was found to correspond to a supercluster of galaxies detected at $z = 0.14$  [paper VI]. 

\subsection{Next steps}

One of the most intriguing
(thus exciting) points raised by our 2016 results is the mismatch between the observed cluster counts and the cosmological predictions from the CMB cosmology [paper II]. A similar problem had independently been pointed out by the Planck cluster counts, but for a much higher mass range and for scaling relations derived in a totally different manner. We are thus facing a dilemma: either there is something that we do not understand in the physics of cluster formation and evolution, or the cosmological model is different or more complicated than currently assumed. We shall use the complete cluster catalogue to investigate this question more in depth.  The enlarged statistical sample will allow us to test the impact of various hypotheses like the ratio $R_{500}/R_{c}$ that was held fixed to 0.15 in our analysis and to proceed with the simultaneous modelling of cosmology, selection effects and cluster evolution. We shall also benefit from the deep high-quality optical coverage of the XXL-N field by the Hyper-Suprime-Camera on the Subaru telescope (HSC Wide Survey\footnote{http://hsc.mtk.nao.ac.jp/ssp/}), which will greatly improve the lensing determination of our cluster masses. A second data release at greater depth will occur in 2017 along with associated scientific articles. We foresee the final data release, including the cluster selection function, for the end of 2018. 

\subsubsection{The X-ray background}

\begin{figure}
\includegraphics[width=8cm]{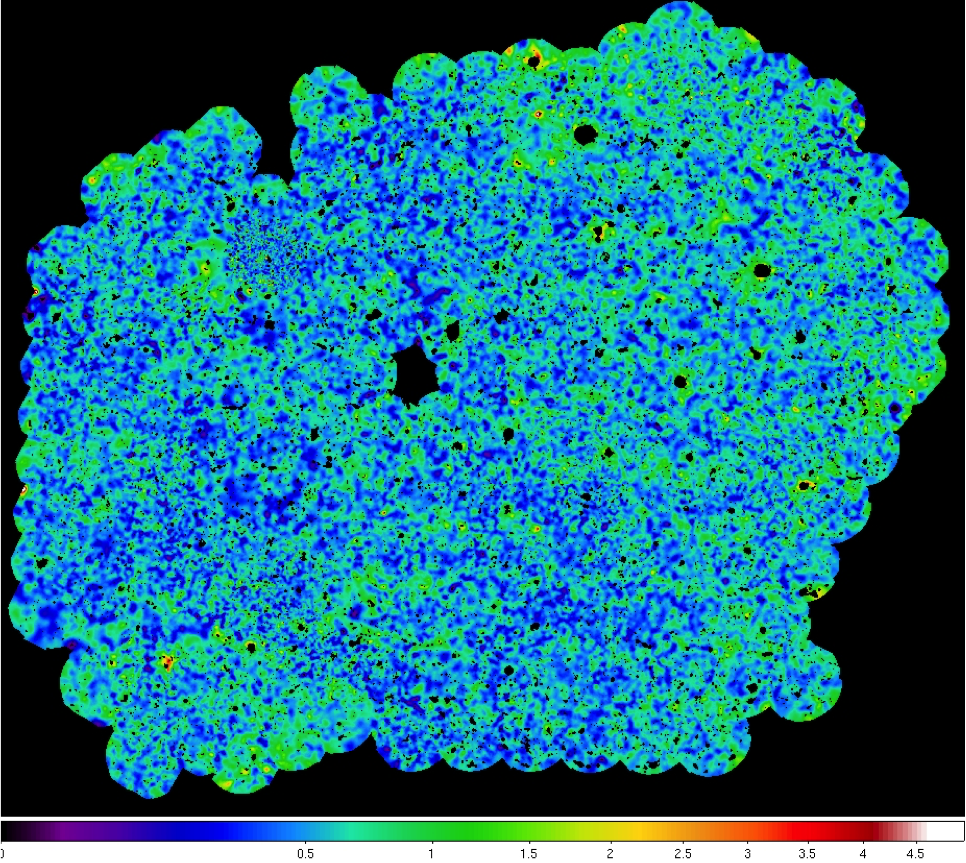}
\caption{View of the X-ray background in the XXL-S field; each circle corresponds to an XMM observation (field of view: $30'$).  The X-ray sources have been removed and the soft and particle backgrounds subtracted. The image is exposure corrected and adaptively  smoothed: some large scale structure is obvious. Covered area: 25 \dd ; displayed band: [0.4-1.3] keV. The colour scale is in unit of counts/s/\dd\ }

\label{xrb}
\end{figure}

The XXL survey enables for the first time the study of the diffuse X-ray background (XRB) on large scales at a high angular resolution and high sensitivity (last studies were on the ROSAT All-Sky Survey data). A first impression of the scientific potential of the X-ray data is rendered by Fig. \ref{xrb}. We are currently working on the characterisation of the structures remaining after source extraction. We are undertaking an auto-correlation study of the map pixels as well as correlations between the X-ray and various maps (HI, IR, FIR) and catalogues (optical and IR galaxies).

\subsubsection{Numerical simulations}
\begin{figure}
\vspace{-1cm}
\includegraphics[width=8cm]{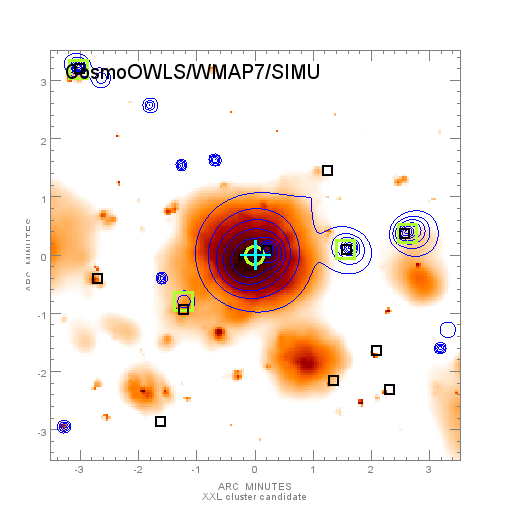}
\includegraphics[width=8cm]{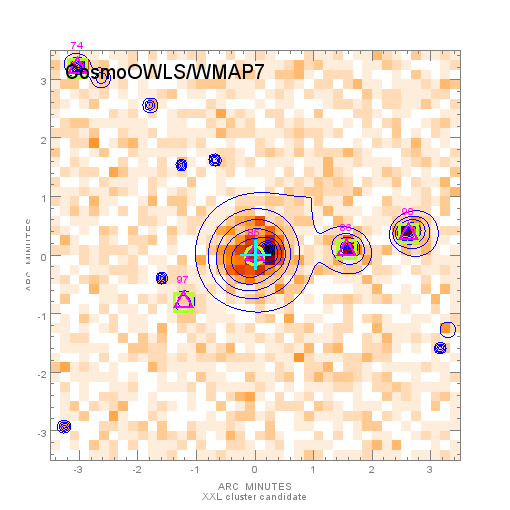}
\caption{Extracted from a cosmo-OWLS lightcone, this simulated 7'x7' image, is centered on a $z=0.95$ cluster having a mass of $M_{500} = 3.5\times 10^{14} M_\odot$ Top: X-ray emissivity map in the  [0.5-2] keV band for the AGN 8.0 model (Le Brun et al 2016 and references therein). The AGN X-ray luminosity is modelled following Koulouridis et al (in prep) and the black squares indicate AGN producing more than 15 photons. Bottom: Corresponding simulated XMM 10 ks image where all instrumental effects are taken into account: PSF, vignetting, diffuse and particle backgrounds. The green circle indicates that the central source has been detected as extended and the green squares stand for point-like sources. }
\label{simul}
\end{figure}

When computing the cluster selection function, we assumed so far that the cluster X-ray emission is spherically symmetric  and follows a $\beta =2/3$ profile; the AGN population was matched to the observed logN-logS, but randomly distributed over the field [paper II]. We shall switch to hydrodynamical simulations, which will provide us with more realistic cluster shapes (mergers, cool cores,  . . .) and with a physical in-situ modelling of the X-ray emission of the AGN population (Koulouridis et al in prep); an example is displayed   in Fig. \ref{simul}. In the end, we shall compute different selection functions, depending on the AGN physics assumed and also on the cosmology. One interesting question is how much the selection function (computed in the flux vs apparent-size plane) is dependent on the assumed cosmology.

\subsubsection{Final cosmological analysis}
The cosmological analysis of the complete cluster sample will be performed using the traditional $dn/dM/dz$ approach. In parallel, we shall use a new method based on X-ray diagnostic diagrams of the cluster population, i.e. relying on observable quantities only: count-rate, hardness-ratio, apparent size, redshift. This allows us to bypass the direct mass determination and thus, to greatly simplify the calculations. Moreover, since we deal with raw X-ray counts, we can include the entire cluster catalogue in the analysis, even those clusters being too faint to estimate their mass (Clerc et al 2012, Pierre et al 2016, submitted).

\section{Prospects for the next decade}
With the new XMM operation mode using 4 reaction wheels, the fuel consumption is halved, which, in principle, will allow the extension of the  XMM observations  up to year $\sim 2028$. A thorough use of this available time will be a matter of trade-off. While there are excellent arguments for undertaking very deep observations of well-defined samples of X-ray emitting objects, there are also compelling reasons to  complete a survey of some 50 \dd\ at a depth of 40ks; let's call it XXL-II. 
Not only will the number of detected objects be significantly higher than that achieved by XXL but, also, the population of currently detected clusters and AGN will be much better characterised. This will have a very noticeable effect on the cosmological analysis (e.g. Pierre et al 2011). In this section, we outline a few key achievements expected from such a deep uniform mosaicking. For that, we assume that the Deep HSC Survey will extend over the entire XXL-N region accordingly (discussion in progress); this will allow highly reliable independent cluster mass measurements.

\subsection{Characterisation of the $z<0.5$ cluster population}
XXL has been very successful in the understanding of the
properties of medium-high mass clusters (T $\geq$ 2keV, i.e. $2\times 10^{13} < M_{500}/M_{\odot} < 10^{14} $ [papers II,III,IV]).
The properties of the lower-mass galaxy group population remain
largely uncharted territory, but is a regime in which XXL-II would have a profound impact. A key question is the degree to
which groups differ from being scaled-down versions of higher mass
clusters, motivated by the expectation that non-gravitational
processes (AGN and SN feedback) are more effective in the group-scale
regime. Recent simulations have shown that scaling relations are best
modelled by an evolving broken power-law (Le Brun et al 2016),
highlighting the decreasing gas fraction as a function of mass [paper XIII]. However, an observational consensus of the presence of a
break in the scaling relations has yet to be reached, with studies
showing the group scaling relations are both consistent
(Sun et al 2009) and inconsistent
(Kettula et al  2015; Lovisari et al 2015) with higher
mass systems. The main drawback of the majority of these works is the small sample sizes and inhomogeneous samples with poorly understood selection biases. XXL-II offers the
opportunity to overcome these drawbacks. At a depth of 40ks, we would be able to measure the temperatures to $\gtrsim$30\% accuracy for all groups out to $z = 0.2$ above $L_{[0.5-2.0]keV}$=10$^{42}$ erg s$^{-1}$  and
out to $z = 0.5$ above $L_{[0.5-2.0]keV}$=5$\times$10$^{42}$ erg s$^{-1}$. This represents the crucial T$\approx$0.5-2 keV range where feedback should dominate over gravity.
The dominance of feedback in low-mass systems leads to large scatter in X-ray luminosity at fixed mass. The amount of, and mass dependence of this scatter are important clues to the nature of the feedback physics. Measuring the scatter can be done by studying clusters  selected through non-ICM properties (e.g. optical tracers). Recent
studies of optically-selected clusters show an increased scatter in
X-ray luminosity compared to X-ray selected samples
(Andreon et al 2016). Indeed, many lower mass groups in the
XXL-N field that are selected from the galaxy and mass assembly (GAMA)
survey are undetected in current XXL data (Giles et al in prep). With the proposed XXL-II, we will be able to measure the full range of $L_{X}$ scatter at a given mass for a complete sample of all GAMA systems with $\geq$10 friends-of-friends members (75 objects).  Moreover, the group mass-range at $z \sim 0.3$, will represent the bulk of the eROSITA sample (see e.g. Borm et al. 2014) but will be observed with an exposure time on average about an order of magnitude lower. XXL-II, with its extensive multi-\LL\ coverage, will provide the multi-band scaling relations that eROSITA will need to fulfill its precision cosmology goal.

\subsection{Census of the $1<z<2$ clusters}
\begin{figure}
%\vspace{14cm}
%\hspace{-4cm}
\center
\includegraphics[width=8cm]{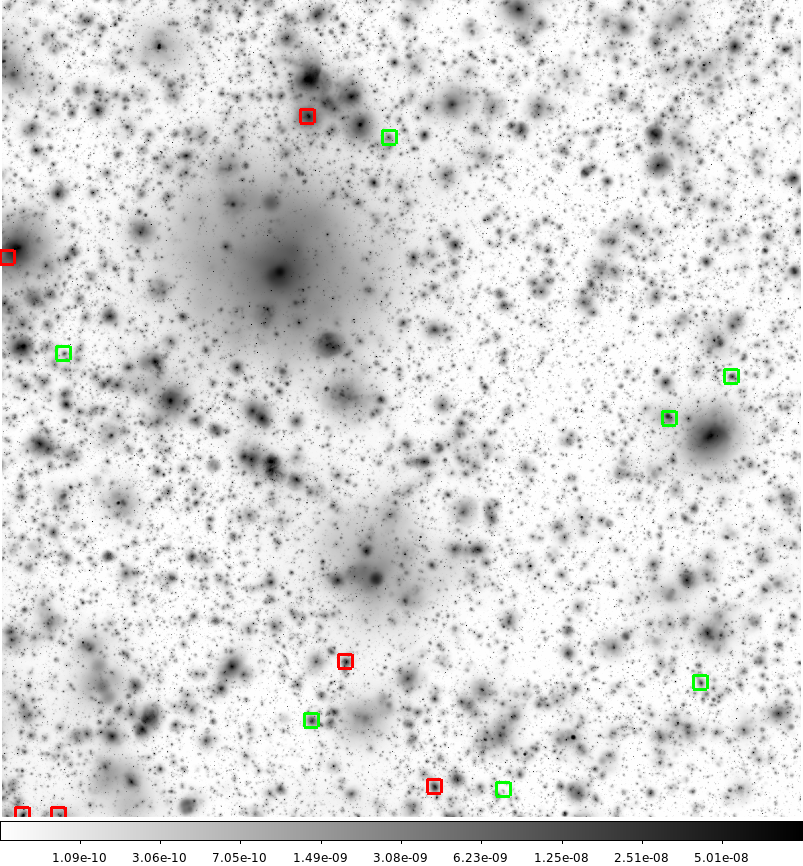}
\includegraphics[width=8cm]{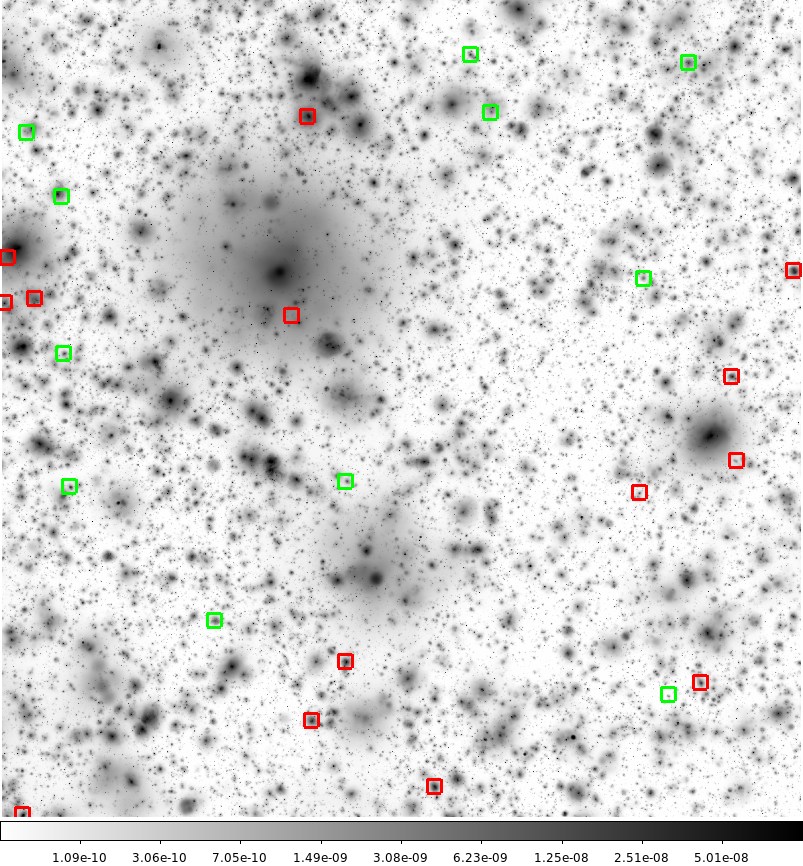}
\caption{Extracted from a cosmo-OWLS AGN 8.0 lightcone, these simulated 3x3 \dd\ X-ray emissivity images show the effect of the XMM sensitivity increase on the detectability of high-redshift clusters. The AGN X-ray emission is modelled in-situ from the black-hole masses and accretion rates given by the simulation.
 The red and green symbols indicate the $z>1$ C1 and C2 detections respectively. Top:  10 ks XXL. Bottom: 40 ks XXL-II. }
\label{highz}
\end{figure}

In the $z>1$ range, we are facing a situation similar to that some 20 years ago, with the Rosat All-Sky Survey (RASS) and and the Palomar Observatory Sky Survey (POSS): clusters around $z=0.4$ were at the sensitivity limit and considered  distant objects. Nowadays, we may replace `$z=0.4$' by `$z=1.2$', `RASS' by `10 ks XMM' and `POSS' by `CFHTLS-Wide'. The difference though, is that we have good reasons to believe, due to comparable advances in numerical simulations, that the  $1<z<2$ range corresponds to the formation epoch of massive clusters and thus, is of extreme cosmological relevance. A few tens of X-ray clusters are known at these distances (and a couple beyond $z>2$, e.g. Gobat et al 2011), but their space density is still undetermined because of the very heterogeneous conditions under which these detections were made. While the observed cluster evolution out to $z \sim 1$ is compatible with self-similarity, there are hints that clusters are fainter at higher redshifts. Our preliminary processing of the cosmo-OWLS AGN 8.0 simulations, duly including the X-ray AGN emission, indicates that we would detect a dozen  $z>1$ C1 clusters (over 50 \dd ) with 10 ks exposures  for the  WMAP7 cosmology. For the Planck 2014 cosmology, the number of high-$z$ detections is doubled; pushing to 40ks exposures would again double the number of detections. Finally, considering the fainter C2 population would add another factor of two. We should then end up with a homogeneous sample of 50-100 $z>1$ clusters, depending on the cosmology and cluster evolution rate\footnote{For the definition of the C1 and C2 cluster selection criteria, refer to [paper II]}.  A visual impression of the sensitivity improvement is given by Fig. \ref{highz}. The gain expected from XXL-II is many-fold: (1) determine the density of high-redshift clusters due to the even X-ray exposure; (2) compare with that from NIR observations, which tends to be much higher, and address the challenging issue of projection effects in galaxy-density  based cluster searches at high-$z$; (3) determine the properties of these objects given the extensive multi-\LL\ coverage - accordingly deep Chandra follow-up would be extremely useful to characterise the AGN population in distant clusters  ; (4) perform a standalone cosmological analysis based on rare-events statistics for the $1<z<2$ range. We note that the systematic exploration of this high-redshift universe at the XXL-II depth is out of the reach of the eROSITA wide survey.

\subsection{AGN}
Under the assumption that the clustering strength of X-ray sources is  independent of the survey flux-limit, then the increase by a factor of two to three of the number of sources in the 40 ksec survey (eg., Cappellutti
et al 2009) could decrease the quasi-Poissonian uncertainties of the correlation function by a factor of at least $\sim$ 4, since: $\sigma_w(\theta)\simeq \sqrt{1+w(\theta)}/DD$,
with DD the number of source pairs within separations $\theta\pm\delta\theta$. However, this could be a rather optimistic reduction of the uncertainties since there is a known dependence of the X-ray source clustering amplitude to the survey flux-limit (Plionis et al. 2008), with lower flux-limited samples showing weaker clustering. In addition, the number of moderate/high-redshift obscured AGN will significantly increase from a deeper exposure in XXL. Furthermore, the X-ray spectral characterization of AGN, currently limited to the brightest sources (largely dominated by unobscured/moderately obscured AGN), will largely improve, allowing for more sophisticated and physically motivated models to be adopted.

\subsection{The X-ray background}
Increasing the exposure of the XXL survey from an average of 10 ks
to 40 ks will have several significant benefits for studies of the cosmic
XRB in addition to the improvement in statistics
(observations of the XRB are nearly always photon limited). Longer
exposures and multiple passes greatly enhance the ability to identify
soft-proton flaring events and either the rejection or modelling and
subtraction of their contribution from images.  Additional exposure also
enables the improved modelling and subtraction of the quiescent particle
background.  Both of these improvements lead to a significant increase
in the reliability of the data.  They are critical due to the relative
faintness of the XRB and scientific relevance of the enabled studies, for example,
the search for the cosmic web.  The increase in statistics will also be
important as the size of useful resolution elements will be decreased
by a factor of two enabling the search for finer structure in the XRB.

\section{Conclusion}

Almost two decades of XMM and Chandra observations have revolutionised much of our knowledge of clusters of galaxies. Moreover, X-ray survey analyses taught us how to handle the many issues impinging on cluster precision cosmology (selection effects, covariance between observables, mass determination and evolutionary physics). While the  publication of the final results of the 10 ks XXL survey will occur in two years time, we propose to start increasing its depth by a factor of four. Given the already existing XMM observations, the total net XMM time to reach a uniform coverage of 40 ks over the 50 \dd\ XXL area is of the order of 13Ms (45 \dd $\times$ 9 pointings/\dd $\times$ 30 ks). This can be easily accommodated at a rate of 2-3 Ms over 6 years, knowing that the  total available Open Time is $\sim$ 15Ms /year.   The main goal is to derive competitive standalone cosmological constraints from the clusters and AGN present in these particular two areas.  Furthermore, the global merit of the project will be greatly enhanced thanks to the synergy between the many associated surveys, from UV to radio. With the new very sensitive instruments such as the HSC in the optical and NIKA2 in the S-Z domains, the scientific potential of the data set will serve a very large scientific community. In the same spirit, we advocate the opening of joint XMM-Chandra Very Large Programmes: along with hydrodynamical simulations, this will definitively enlighten the physics and evolution of the low-mass $z \sim 0.5$ and high-redshift clusters, in relation to galactic nucleus activity. XXL-II will bridge the gap between the expected eROSITA and Athena performances in terms of combined  sensitivity, coverage and angular resolution. It will open a totally new  field for XRB research and will constitute a unique legacy for the next generations, particularly for the cosmological exploitation of the eROSITA and Euclid missions.

\section{Affiliations}

\begin{itemize}
\item[$^{1}$] Service d'Astrophysique du CEA, Saclay, FR                	%1
\item[$^{2}$] Laboratoire d'Astrophysique, Marseille, FR                      	%2
\item[$^{3}$] University of Bristol, UK                                        %3
\item[$^{4}$] INAF-IASF, Milano, IT                                   	%4
\item[$^{5}$] INAF-OABO, Bologna, IT                                 	%5
\item[$^{6}$] University of Michigan, Ann Arbor, US                  	%6
\item[$^{7}$] Chalmers University of Technology, Onsala, SE           	%7
\item[$^{8}$] INAF-OAB, Brera, IT                                        	%8
\item[$^{9}$] Australian Astronomical Observatory, Epping, AU          	%9
\item[$^{10}$] Observatoire de la Cote d'Azur, Nice, FR                 	%10
\item[$^{11}$] University of Liverpool, UK                               	%11
\item[$^{12}$] NAOJ, Tokyo, JP                                               	%12
\item[$^{13}$] Argelander-Institut fur Astronomie, Bonn, DE             	%13
\item[$^{14}$] ISDC, Geneva Observatory, CH                           	%14
\item[$^{15}$] Aristotle University, Thessaloniki, GR         		%15
\item[$^{16}$] Ulugh Beg Astronomical Institute, Tashkent, UZ           	%16
\item[$^{17}$] University of Zagreb, HR			 	  	%17
\item[$^{18}$] NASA, GSFC, US                                         	%18
\item[$^{19}$] University of Li\`ege, BE                                  	%19
\item[$^{20}$] University of Victoria, CA                             	%20
\item[$^{21}$] Istanbul University,TR                       %21
\item[$^{22}$] European Space Astronomy Center, Madrid, ES            	%22
\item[$^{23}$] INAF-OAR, Roma, IT                                               %23
\item[$^{24}$] University of Western Austalia, AU                     	%24
\item[$^{25}$] Observatoire de Paris, FR                              	%25
\item[$^{26}$] MPI for Extraterrestrial Physics, Garching, DE 	%26
\item[$^{27}$] IIT Hyderabad, IN                                	%27
\item[$^{28}$] University of Oxford, UK                                 	%28
\item[$^{29}$] Uppsala University, SE                  	%29
\item[$^{30}$] H. S. Center for Astrophysics, Cambridge, US              	%30
\item[$^{31}$] Observatory of Athens, GR                                	%31
\item[$^{32}$] INAF-OAP, Padova, IT                                      	%32
\item[$^{33}$] Service d'Informatique du CEA, Saclay, FR                    	%33
\item[$^{34}$] University of Chicago, US                                 	%34
\item[$^{35}$] Leiden University, NL                                     	%35
\item[$^{36}$] Service de Physique des Particules du CEA, Saclay, FR       	%36
\item[$^{37}$] European Southern Observatory, Garching, DE              	%37
\item[$^{38}$] University of Birmingham, UK                              	%39
\item[$^{39}$] University of Lund, SE % 38
\item[$^{40}$] University of Colorado, Boulder, US                                	%40
\item[$^{41}$] ASIAA, Taipei, TW                                        	%41
\item[$^{42}$] Institut de Physique Theorique du CEA, Saclay, FR            	%42
\item[$^{43}$] University Paris-Diderot, Paris, FR                     	%43
\item[$^{44}$] Jodrell Bank, Manchester, UK                          %44
\item[$^{45}$] University La Sapienza, Rome, IT                     	%45
\item[$^{46}$] Institut d'Astrophysique de Paris, FR                        	%46
\item[$^{47}$] Yamagata University, JP %47
\item[$^{48}$] Victoria University, Wellington, NZ %48
\item[$^{49}$] University Roma Tre, Rome, IT %49
\item[$^{50}$] Hiroshima University, JP %50
\item[$^{51}$] York University, Toronto, CA %51
\item[$^{52}$] IPAC, Pasadena, US %52
\item[$^{53}$] Arcetri Observatory, Florence, IT %53
\item[$^{54}$] University of Sussex, Brighton, UK %55
\item[$^{55}$] University of Tor Vegata, Roma, IT %56
\item[$^{56}$] University of Bologna, IT
\end{itemize}


\begin{thebibliography}{}
  \bibitem{} XXL paper I: Pierre M. et al 2016, A\&A 592, A1
  \bibitem{} XXL paper II: Pacaud F. et al 2016, A\&A 592, A2
   \bibitem{} XXL paper III: Giles P. et al 2016, A\&A 592, A3
    \bibitem{} XXL paper IV: Lieu M. et al 2016, A\&A 592, A4
     \bibitem{} XXL paper V: Mantz A. et al 2014, ApJ 794, 157
  \bibitem{} XXL paper VI : Fotopoulou et al 2016, A\&A 592, A6   
    \bibitem{} XXL paper VII: Pompei E.et al 2016 A\&A, 592, A7
     \bibitem{} XXL paper XIII: Eckert D. et al 2016, A\&A, 592, A2 
\bibitem{} Allen et al 2011, ARA\&A 49, 409     
\bibitem{} Andreon et al 2016, A\&A 585, 147    
 \bibitem{} Borm et al 2014, A\&A 567, 65
\bibitem{} Cappelluti et al 2009, A\&A 497, 635
 \bibitem{} Chiappetti et al 2013, MNRAS 429, 1652
 \bibitem{} Clerc N. et al 2014,  MNRAS 423, 356
 \bibitem{} Gobat R et al 2011, A\&A 526, 133 
\bibitem{} Elyiv et al 2012,  A\&A 537, 131
  \bibitem{} Kettula K. et al 2015, MNRAS 451, 1460
\bibitem{} Le Brun A. et al   2016, arXiv:1606.04545
  \bibitem{} Lovisari L. et al 2015, A\&A 573, 118
\bibitem{} Pierre M.  et al 2011, MNRAS 414, 1732
\bibitem{} Plionis, M. et al., 2008, ApJ, 674, L5
  \bibitem{} Sun et al 2009, ApJ 693, 1142
\bibitem{} \underline{More information} can be found on the website of:\\ {\sl Hot Spots in the XMM Sky: Cosmology from X-ray to radio} \\
a prospective conference held in Mykonos (June 2016):\\
 http://www.astro.auth.gr/$\sim$xmmcosmo16  
\end{thebibliography}
\end{document}